\documentstyle[emulateapj,apjfonts,psfig]{article}

\lefthead{J.~M.~Miller et al.}  \righthead{XTE J1751$-$305}

\received{Date}
\revised{Date}
\accepted{Date}

\journalid{vol}{date}
\articleid{1}{4}
\paperid{id}

\cpright{AAS}{1999}
\ccc{x}

\begin{document}

\title{\textit{XMM-Newton} Spectroscopy of the Accretion-Driven
Millisecond X-ray Pulsar\\ XTE~J1751-305 in Outburst}

\author{J.~M.~Miller\altaffilmark{1,2,3}, 
	R.~Wijnands\altaffilmark{1,4,5},
	M.~M\'endez\altaffilmark{6},
	E.~Kendziorra\altaffilmark{7}
	A.~Tiengo\altaffilmark{8},
	M.~van~der~Klis\altaffilmark{8},
	D.~Chakrabarty\altaffilmark{1},
	B.~M.~Gaensler\altaffilmark{2},
	and W.~H.~G.~Lewin\altaffilmark{1}
	}

\altaffiltext{1}{Center~for~Space~Research and Department~of~Physics,
        Massachusetts~Institute~of~Technology, Cambridge, MA
        02139--4307}
\altaffiltext{2}{Harvard-Smithsonian Center for Astrophysics, 60
	Garden Street, Cambridge, MA 02138, jmmiller@head-cfa.harvard.edu}
\altaffiltext{3}{National Science Foundation Astronomy and Astrophysics Fellow}
\altaffiltext{4}{School of Physics and Astronomy, University of St. Andrews, St
	Andrews KY16 9SS, UK}
\altaffiltext{5}{Chandra Fellow while at MIT}
\altaffiltext{6}{SRON, National Institute for Space Research, Sorbonnelaan 2,
        NL-3584 CA, Utrecht, NL}
\altaffiltext{7}{Institut f\"ur Astronomie und Astrophysik,
        Abt. Astronomie, Universit\"at T\"ubingen, Sand 1, D-72076,
        T\"ubingen, DE}
\altaffiltext{8}{Astronomical Institute ``Anton Pannekoek'',
	University of Amsterdam, Kruislaan 403, NL-1098 SJ Amsterdam, NL}


\authoremail{jmm@space.mit.edu}

\label{firstpage}

\begin{abstract}
We present an analysis of the first high-resolution spectra measured
from an accretion-driven millisecond X-ray pulsar in outburst.  We
observed XTE J1751--305 with {\it XMM-Newton} on 2002 April 7 for
approximately 35~ks.  Using a simple absorbed blackbody plus power-law
model, we measure an unabsorbed flux of $6.6 \pm 0.1 \times 10^{-10}~
{\rm erg}~ {\rm cm}^{-2}~ {\rm s}^{-1}$~ (0.5--10.0 keV).  A hard
power-law component ($\Gamma = 1.44 \pm 0.01$) contributes 83\% of the
unabsorbed flux in the 0.5--10.0 keV band, but a blackbody component
($kT = 1.05 \pm 0.01$~ keV) is required.  We find no clear evidence
for narrow or broad emission or absorption lines in the time-averaged
spectra, and the sensitivity of this observation has allowed us to set
constraining upper-limits on the strength of important features.  The
lack of line features is at odds with spectra measured from some other
X-ray binaries which share some similarities with XTE J1751--305.  We
discuss the implications of these findings on the accretion flow
geometry in XTE~J1751--305.
\end{abstract}

\section{Introduction}
Millisecond radio pulsars are thought to be created in neutron star
low-mass X-ray binaries (LMXBs). In those LMXBs, accreting matter may
spin-up the neutron star (see, e.g., Bhattacharya \& van den Heuvel
1991).  Therefore, it is expected that millisecond pulsars should also
be found in LMXBs during the accretion phase of the binary as X-ray
pulsars.  Although evidence for rapidly spinning neutron stars in
LMXBs was inferred from the burst oscillations which were seen during
type-I X-ray bursts in several systems (see Strohmayer 2001 for a
review), the detection of millisecond pulsations in persistent
emission remained elusive for many years.  In 1998, the first such
system was discovered (SAX J1808.4--3658, which has a spin frequency
of 401 Hz; Wijnands \& van der Klis 1998a).  This source was
extensively studied due to its obvious importance for binary evolution
scenarios and accretion flow geometries and dynamics (see, e.g.,
Wijnands \& van der Klis 1998b and references therein).  

In the spring of 2002, Markwardt \& Swank (2002a) reported the
discovery of the second accretion-driven millisecond pulsar XTE
J1751--305 (435 Hz) in the {\it Rossi X-ray Timing Explorer} ({\it
RXTE}) bulge scan observation program.  About one month after the
discovery of this system, XTE J0929--314 was discovered (185 Hz;
Remillard, Swank, \& Strohmayer 2002.).  The neutron stars in both
systems are in orbit around a companion star, with an orbital period
of $\sim$42 minutes (Markwardt \& Swank 2002b; Galloway et al. 2002).
These systems are very tight binaries and the inferred mass of their
companion stars is very low ($\sim 0.01~M_{\odot}$).

After the discovery of XTE J1751--305, we submitted a
target-of-opportunity request to {\it XMM-Newton} for the purpose of
studying this source with high-resolution spectroscopy.  The X-ray
spectrum of the first system --- SAX J1808.4--3658 --- could only be
studied during outburst using {\it RXTE} and with the Wide Field
Cameras aboard\textit{BeppoSAX}, which have only moderate spectral
resolution.  In this {\it Letter}, we present an analysis of the
time-averaged EPIC-pn and Reflection Grating Spectrometer (RGS) data.
The spectra obtained represent the first CCD- and gratings-resolution
measurements from a millisecond X-ray pulsar in outburst.

\section{Observation and Data Reduction}
\textit{XMM-Newton} observed XTE~J1751$-$305 beginning on 2002 April
07.5 (UT).  The EPIC-pn camera was operated in ``timing'' mode to
prevent photon pile-up effects and to allow pulse-phase-resolved
spectroscopy .  The RGS was operated in the standard ``spectroscopy''
mode.  The ``thin'' optical blocking filter was selected for the EPIC
cameras.  The EPIC-MOS cameras were not operated in modes optimized
for a source of this intensity, and we do not consider the MOS spectra
here.  

The EPIC-pn and RGS data were reduced using SAS version 5.3.  We
applied the standard reduction procedure ``epproc'' to produce a
calibrated pn event list.  In ``timing'' mode, the spatial information
is compressed into one dimension.  We extracted source data in a
rectangle (with a width of 37'') along the length of the CCD, and
background data from adjacent regions; only ``single'' and ``double''
events were selected.  The net pn exposure was 33.7~ks.  Using
LHEASOFT version 5.1, the tool ``grppha'' was used to rebin the
spectrum to require a minimum of 20 counts per bin.  We used the trial
``timing'' mode redistribution and ancillary response functions
developed by the EPIC-pn calibration team.  We applied the standard
reduction procedure ``rgsproc'' to produce calibrated RGS event lists,
first-order spectra, and response functions.  Periods of high
instrumental background were excluded, giving a net RGS exposure of
34.2~ks.  The RGS spectra were grouped to require at least 20 counts
per bin.

\section{Analysis and Results}
Preliminary spectroscopic results from this {\it XMM-Newton}
observation were reported in Miller et al. 2002a and Miller et
al. 2002b.  The accurate position of XTE J1751--305 was reported by
Ehle et al. (2002), also based on analysis of this observation (${\rm
R.A.} = 17h51m13.5s, {\rm Decl.} = -30^{\circ} 37'22''$, equinox
2000.0).

The internal calibration accuracy of the EPIC cameras is better than
5\% for on-axis sources (Kirsch 2002).  We find deviations up to
0.6~keV which are not removed by spectral modeling, and therefore
restrict our analysis of the pn spectrum to the 0.6--10.0~keV band.
The average count rate (79.36 counts~${\rm s}^{-1}$) is well below
that at which photon pile-up occurs in timing mode (1500 counts~${\rm
s}^{-1}$).  Tests of the effective area calibration of the RGS reveal
accuracies of 5\% in the 7--36\AA~ (0.35--1.77~keV) range and 10\% at
wavelengths shorter than 7\AA~ (energies above 1.77~keV; den Herder
2002).  Although our best-fit model for the EPIC-pn spectrum (see
below) provides an adequate fit to the RGS spectrum ($\chi^{2}/\nu =
1.314, \nu = 1463$), broad deviations are apparent at energies above
1.8~keV.  We therefore rely on the EPIC-pn spectrum to characterize
the continuum emission, and search the RGS band in contiguous 3\AA~
slices for narrow features.

The EPIC-pn and RGS spectra were analyzed using XSPEC version 11.1.
All errors quoted in this paper are 90\% confidence errors.  Note that
systematic errors are not added to the spectra to account for flux
calibration uncertainties.

\subsection{The EPIC-pn Spectrum}
We first considered a model consisting of blackbody and power-law
components, modified by neutral absorption in the ISM (using the
``phabs'' model in XSPEC).  We measure a best-fit equivalent hydrogen
column density of $N_{H} = (9.8\pm 0.1) \times 10^{21}~ {\rm atoms}~
{\rm cm}^{-2}$.  The blackbody temperature obtained is moderate: $kT =
1.05 \pm 0.01$~keV.  For spherical symmetry, this translates into a
blackbody radius of $R = f^{2} \times 3.01^{+0.09}_{-0.06}~ (d /
10~{\rm kpc})~ {\rm km}$, where the spectral hardening factor $f$ is
the ratio of the color temperature and the effective temperature, and
$d$ is the distance to the source.  The measured power-law index is
$\Gamma = 1.44\pm 0.01$; the normalization is $(6.4 \pm 0.1) \times
10^{-2}$ (${\rm photons}~ {\rm keV}^{-1}~ {\rm cm}^{-2}~ {\rm s}^{-1}$
at 1 keV).  This model produces an adequate fit to the data:
$\chi^{2}/\nu = 1.107$ (where $\nu$ is the number of degrees of
freedom; $\nu = 1884$ for this model; see Figure 1).  We note that a
simple power-law model returns a very poor fit ($\chi^{2}/\nu > 4, \nu
= 1886$); the soft blackbody component is strongly required by the
data.

With this simple but standard model, we obtain an unabsorbed
0.5--10.0~keV flux of $(6.6 \pm 0.1) \times 10^{-10}~ {\rm erg}~ {\rm
cm}^{2}~ {\rm s}^{-1}$, or $(0.170 \pm 0.004)~ {\rm photons}~ {\rm
cm}^{-2}~ {\rm s}^{-1}$.  The hard power-law component contributes
83\% of the unabsorbed flux 

\centerline{~\psfig{file=fig1.epsi,width=3.5in,angle=-90}~}
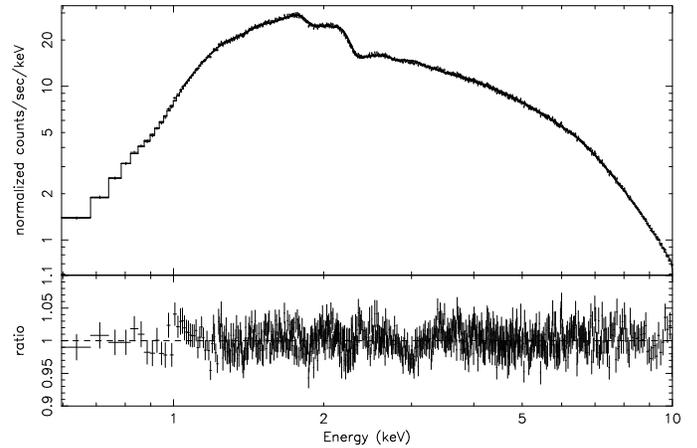
\figcaption[t]{\small The EPIC-pn spectrum of XTE~J1751$-$305, fit
with a model consisting of blackbody and power-law components modified
by photoelectric absorption (see Section 3 for details).  The spectrum
and data/model ratio shown above are rebinned for visual clarity.}
\medskip

in the 0.5--10.0~keV band (an extrapolation from the 0.6--10.0~keV
band as the 0.5--10.0~keV range is more conventional).

Motivated by evidence for a Comptonizing volume in SAX~J1808.4--3658
(Gierlinski, Done, \& Barret 2002), we also fit the EPIC-pn spectrum
with the ``compTT'' Comptonization model (Titarchuk 1994), which
attempts to account for the Compton-upscattering of cool photons in a
hot corona self-consistently.  This model does not provide an
acceptable fit by itself ($\chi^{2}/\nu = 2-4, \nu = 1884$).  We find
a degeneracy between a cool corona ($kT \sim 3$~keV) with high optical
depth ($\tau \sim 8$), and a hot corona ($kT \sim 40$~keV) with lower
optical depth ($\tau \sim 1$), each up-scattering a seed photon
distribution peaking between $kT =$0.5--0.6~keV.  Finally, we tried
fitting a blackbody component in addition to compTT.  This model might
correspond to a scenario in which the blackbody emission region is
only partially covered by the corona.  This model also yielded a poor
fit ($\chi^{2}/\nu \sim 2$).

The strength, ionization state, and profile of Fe~K$\alpha$ emission
lines --- if produced through irradiation of the accretion disk ---
can reveal a great deal about the accretion geometry of low-mass X-ray
binaries.  Asai et al. (2000) report the possible detection of an
Fe~K$\alpha$ line in 4U 1820--30 (P$_{orb} = 11.4$~minutes; Stella,
White, \& Priedhorsky 1987).  The line was found to be ionized (E $=
6.6 \pm 0.1$~keV), broad (FWHM $=0.7^{+0.2}_{-0.5}$~keV), and weak
($W_{K\alpha} = 33^{+12}_{-11}$~eV).  Assuming a line of equivalent
FWHM in the 6.40--6.97~keV range (Fe~I --- Fe~XXVI), we measure a 95\%
confidence upper-limit of just 6~eV in the EPIC-pn spectrum of XTE
J1751--305.  Using a Gaussian with zero width to model a line narrower
than the EPIC-pn resolution, the 95\% confidence upper limit on the
equivalent width of any such narrow Fe~K$\alpha$ lines is less than
4~eV (6.40--6.97~keV).

\subsection{Pulse-Phase-Resolved Spectroscopy: First Results}
We applied the SAS task ``barycen'', and using the binary system
parameters reported by Markwardt et al. (2002) we produced spectra
from the ``high'' and ``low'' parts of the pulse.  We find that the
model applied to the time-averaged spectrum is an acceptable
description of these spectra (no convincing narrow or broad emission
or absorption features are found).  Interestingly, the slope of the
power-law component is consistent in the two spectra but the
normalization changes, and the blackbody component changes in
temperature but not in normalization.  

\centerline{~\psfig{file=fig2.epsi,width=3.5in,angle=-90}~}
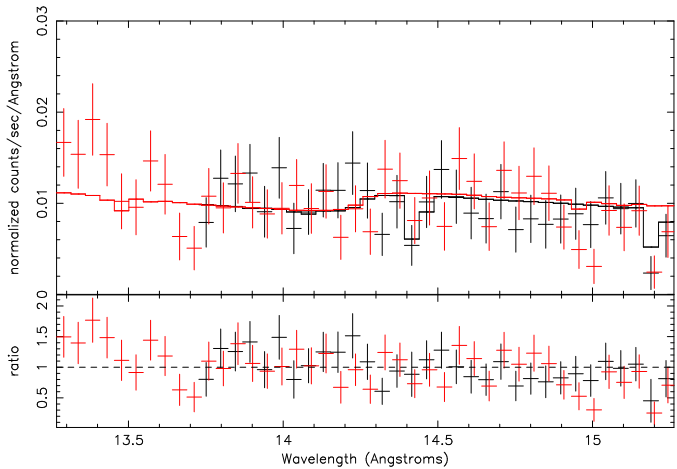
\figcaption[t]{\small The RGS spectra of XTE~J1751$-$305 (RGS-1 in
black, RGS-2 in red; both are rebinned for clarity and narrow features
are instrumental) fit locally in the region of the Ne K edge
(14.25\AA) with a simple power-law plus edge model.  The edge depth
was fixed at the solar value but the power-law index and normalization
were allowed to vary; a trend in the data/model ratio is clearly
visible.  Fits made with a variable edge depth suggest Ne is likely
under-abundant (see text).}
\medskip

Similar behavior was discovered in pulse-phase-resolved spectra of SAX
J1808.4$-$3658 (Gierlinski, Done, \& Barret 2002).

We note that a possible feature at approximately 2.9~keV appears to
vary in width, intensity, and centroid energy in the ``high'' and
``low'' spectra; however, the error limits on the line parameters
overlap.  This feature is also visible in the time-averaged spectrum
(see Figure 1).  As there are no astrophysically abundant elements
with transitions near this energy, it is tempting to associate this
feature with a red-shifted absorption line from the neutron star
surface (given the temperature of the blackbody, perhaps Fe~XXV or
Fe~XXVI).  Although an F-test finds that the addition of a Gaussian to
model this feature is significant at the 7$\sigma$ level of confidence
in the ``low'' part of the pulse (3$\sigma$ in the ``high'' part), a
similar feature is apparent when the EPIC-pn spectrum of the Galactic
black hole candidate XTE~J1650$-$500 (Miller et al. 2002c) is fit
using the same response matrix.  It is likely that this feature is due
to a defect in the timing-mode response matrix.

\subsection{The RGS Spectra}
We analyzed the RGS spectra in contiguous 3\AA~ slices for narrow
emission or absorption lines.  In each 3\AA~ slice, a simple power-law
plus ISM edge(s) model was fit to the data.  We find no convincing
evidence for narrow absorption features in this band (0.3--2.5~keV, or
5--40\AA).  At energies below $\sim$0.6~keV (wavelengths longer than
$\sim$21\AA) the sensitivity is low due to the relatively high
column density along the line of sight to XTE J1751--305.

Recently, evidence for enhanced Ne abundances has been found in
compact binaries (Schulz et al. 2001; Juett, Psaltis, \& Chakrabarty
2001).  We fit the RGS spectra around the position of the Ne K edge
(14.25\AA~) with a power-law model between 13.25\AA~ and 15.25\AA (see
Figure 2).  The power-law index and normalization were both allowed to
vary in this fit.  Using the cross-sections of Verner et al. (1993)
and solar abundances relative to H as per Morrison \& McCammon (1983),
we find that Ne may be {\it under}-abundant in XTE J1751--305: 

\centerline{~\psfig{file=fig3.epsi,width=3.5in,angle=-90}~}
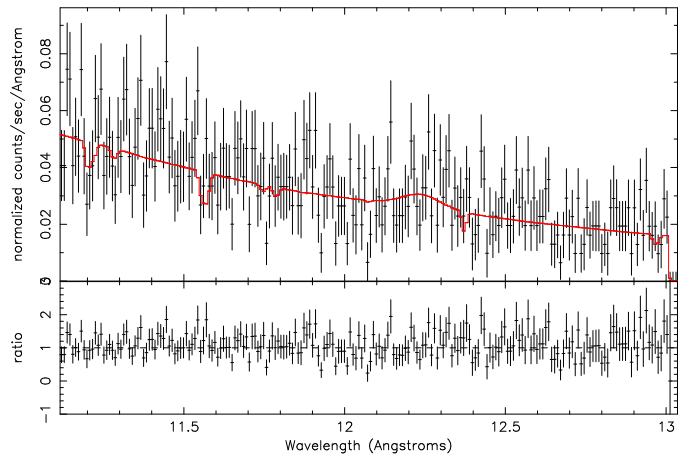
\figcaption[t]{\small The RGS-2 spectrum of XTE~J1751$-$305 in the
region of the Ne~X Ly-$\alpha$ line at 12.13\AA, fit with a simple
power-law plus Gaussian model.  Our best fit model does not require an
emission line within $12.13\pm 0.1$\AA, assuming a FWHM equivalent to
the prominent Ne~X Ly-$\alpha$ emission line seen in the 42-minute
X-ray pulsar 4U 1626--67 (Schulz et al. 2001).  Narrow features are
instrumental.}
\medskip

the 95\% confidence upper-limit is that the abundance of Ne is only
77\% of the solar value.  A solar Ne abundance does not allow for an
optimal characterization of the local spectrum (see Figure 2).

The binary X-ray pulsar 4U 1626--67 also has a 42-minute orbital
period.  Moreover, the distance to this system may be close to 8~kpc
(Chakrabarty 1998); we anticipate a Galactic center location for
XTE~J1751$-$305 and therefore a similar distance.  A {\it
Chandra}/HETGS spectrum of 4U~1626$-$67 reveals broadened emission
lines and Doppler-shifted line pairs (Schulz et al. 2001).  The Ne
Ly-$\alpha$ line (12.13\AA) is particularly strong (a flux of $8.15
\pm 0.93 \times 10^{-5}~ {\rm photons}~ {\rm cm}^{-2}~ {\rm s}^{-1}$
is measured; FWHM $=2860 \pm 330~ {\rm km}~ {\rm s}^{-1}$).  We report
the 95\% confidence upper-limits on emission lines from helium-like
and hydrogenic Ne, Mg, and Si in Table 1 (upper limits on O emission
lines are not constraining due to poor statistics at low energies).

\section{Discussion}
We have analyzed the first CCD- and grating-resolution X-ray spectra
of a millisecond X-ray pulsar in outburst.  We find no convincing
evidence for broad or narrow emission or absorption features in the
the EPIC-pn and RGS spectra (see Figure 1 and Table 1).  With our
simple blackbody plus power-law model for the EPIC-pn spectrum, we
find that the power-law component comprises 83\% of the unabsorbed
flux in the 0.5--10.0~keV band.  The power-law component is hard:
$\Gamma = 1.44 \pm 0.01$, but within a range typical for X-ray pulsars
(e.g., $\Gamma =$ 1.0--1.5; White, Swank, \& Holt 1983).  The EPIC-pn
spectrum requires a blackbody with $kT = 1.05 \pm 0.01$~keV and an
implied emitting radius of $R = f^{2} \times 3.01^{+0.09}_{-0.06}~
(d/10 {\rm kpc})$~km.  It is fair to say that our knowledge of the
spectral hardening factor $f$ is rather poor (for a discussion, see
Lewin, van Paradijs, \& Taam 1993).  If $f$ is less than 1.3, it would
indicate that only part of the neutron star surface (a ``hot spot'')
is involved.  

Markwardt et al. (2002) fit a simple absorbed power-law model to the
{\it RXTE} spectrum of XTE~J1751--305, and measure a softer power-law
index ($1.7 < \Gamma < 1.9$).  The effective lower sensitivity bound
of the {\it RXTE} Proportional Counter Array (PCA) is 3 keV; fitting
the 3--10~keV EPIC-pn spectrum with only a power-law, we find $\Gamma
= 1.72 \pm 0.01$ for $N_{H} = (9.8\pm 0.1) \times 10^{21}~ {\rm
atoms}~{\rm cm}^{-2}$ and $\Gamma = 1.90 \pm 0.02$ allowing $N_{H}$ to
float (in this case, $N_{H} = [2.5\pm 0.1] \times 10^{22}~ {\rm
atoms}~{\rm cm}^{-2}$).  Thus, the discrepant indices are easily
explained in terms of different instrumental ranges.

In modeling simultaneous {\it Chandra}/LETGS and {\it RXTE} spectra of
XTE J0929--314, Juett, Galloway, \& Chakrabarty (2002) measure
similarly discrepant power-law indices.  They suggest an astrophysical
origin for the discrepancy in XTE J0929--314 and XTE~J1751--305 in the
form of a power-law with a break or a roll-over in the 1.4--4.4~keV
range.  A model consisting of blackbody and broken power-law
components ($E_{break} = 3.7 \pm 0.1~{\rm keV}x,~ \Gamma_{E < 3.7} =
1.34 \pm 0.03,~ \Gamma_{E > 3.7} = 1.60 \pm 0.04$) provides an
improved fit to the spectrum of XTE J1751--305 ($\chi^{2}/\nu = 1.090,
\nu = 1882$).  Although these results suggest that an astrophysical
origin for the apparent spectral evolution is possible, a number of
concerns remain.  Given that: (1) the effective area (flux)
calibration of the EPIC-pn has 5\% uncertainties (Kirsch 2002), (2)
the LETGS (plus ACIS-S array) flux calibration has 10\% uncertainties
above 1~keV (H. Marshall, priv. comm.), (3) that {\it RXTE} PCA and
High Energy X-Ray Timing Experiment (HEXTE) observations of the
``Crab'' nebula (taken to be a pure power-law for calibration) differ
in power-law index by $\delta(\Gamma) \simeq 0.1$ (see, e.g., Wilms et
al. 1999), and (4) that Markwardt et al. (2002) did not fit a
blackbody component to the {\it RXTE} spectra of XTE~J1751--305, we do
not consider an astrophysical origin for the discrepant indices to be
required by the present data.

The upper limits on the strength of narrow (consistent with instrument
resolution; FWHM $\sim$ 0.1~keV) and broad (FHWM $=$ 0.7~keV) Fe
K$\alpha$ emission lines are 4~eV and 6~eV, respectively (95\%
confidence).  This stands in contrast the detection of a weak Fe
K$\alpha$ line in outburst spectra of the 401~Hz millisecond X-ray
pulsar SAX J1808.4--3658 obtained with {\it RXTE} (Heindl \& Smith
1998; Gierlinski, Done, \& Barret 2002).  It is possible that a
strong, hot corona (or the magnetosphere, if the coranae in these
LMXBs is small or disrupted) which produces to the power-law component
may ionize the disk in XTE~J1751$-$305 to a degree which prevents
Fe~K$\alpha$ emission.

The absence of H-like and He-like resonance emission lines like those
observed in 4U 1626--67 (Schulz et al. 2001) in the spectrum of XTE
J1751--305 is also consistent with a highly ionized disk in this
source.  Secondarily, the absence of such lines may be due to
differences in the accretion flow geometry due to the neutron star
magnetic field.  Whereas the magnetic field in XTE~J1751$-$305 is
likely to be similar to that in SAX~J1808.6$-$3658 ($B = (2-6) \times
10^{8}$~G; Wijnands \& van der Klis 1998a), the magnetic field in
4U~1626$-$67 is likely much higher ($B = 3 \times 10^{12}$~G,
Orlandini et al. 1998).  The stronger magnetic field in 4U 1626--67
may disrupt the inner disk, allowing cool material in the outer disk
to be irradiated by a weaker corona.  Emission lines have also been
seen in neutron star systems which are viewed edge-on (so-called
accretion disk coronae or ``ADC'' sources; see, e.g., Kallman et
al. 2002); the absence of emission lines in XTE J1751--305 may also be
partially due to a low system inclination (Markwardt et al. 2002
report no evidence of dips or eclipses).

Studies of some ultra-compact systems suggest a Ne-rich companion (for
observational results, see: Schulz et al. 2001; Juett, Psaltis, \&
Chakrabarty 2001; for theoretical discussions see Yungelson, Nelemans,
\& van den Heuvel 2002; Bildsten 2002).  Fits to the Ne photoelectric
absorption edge in the RGS spectra of XTE J1751--305 constrain the
abundance of Ne to be at most 77\% of the solar value along this line
of sight (95\% confidence upper-limit; see Section 3.3).  This may
suggest that a Ne-rich companion is unlikely in the case of XTE
J1751--305.

\section{Acknowledgments}
We are grateful to the \textit{XMM-Newton} Project Scientist Fred
Jansen for granting time to observe this source.  We thank Michael
Nowak for useful discussions.  J. M. M. is grateful for support from
the NSF.  R. W. was supported by NASA through Chandra fellowship grant
PF9-10010, which is operated by the Smithsonian Astrophysical
Observatory for NASA under contract NAS8-39073.
W. H. G. L. gratefully acknowledges support from NASA.  This work was
supported in part by the Netherlands Organization for Scientific
Research (NWO).  This research has made use of the data and resources
obtained through the HEASARC on-line service, provided by NASA-GSFC.

\begin{table}[h]
\caption{Emission Line Upper-Limits}
\begin{footnotesize}
\begin{center}
\begin{tabular}{lcccc}
 ~ & \multicolumn{2}{c}{Helium-like} & \multicolumn{2}{c}{Hydrogenic}\\
 ~ & Broad$^{a}$ & Narrow$^{b}$ & Broad$^{a}$ & Narrow$^{b}$ \\
 ~ & ($10^{-5}~{\rm ph}~{\rm cm}^{-2}~{\rm s}^{-1}$) & ($10^{-5}~{\rm
 ph}~{\rm cm}^{-2}~{\rm s}^{-1}$) & ($10^{-5}~{\rm ph}~{\rm
 cm}^{-2}~{\rm s}^{-1}$) & ($10^{-5}~{\rm ph}~{\rm cm}^{-2}~{\rm
 s}^{-1}$) \\
\tableline
Ne & 1.6 & 1.4 & 0.9 & 0.5 \\
Mg & 1.4 & 1.3 & 11 & 8.8 \\
Si & 29 & 26 & 3.7 & 3.5 \\
\end{tabular}
\vspace*{\baselineskip}~\\ \end{center} \tablecomments{The 95\%
confidence upper limits on highly ionized resonance emission lines
within 0.1\AA~ of their expected rest-frame wavelengths in
the RGS spectra of XTE J1751--305.  $^{a}$ Assuming FWHM $=$ 3000
km/s, similar to the Doppler-shifted pairs in 4U 1626--67 (Schulz et
al. 2001).  $^{b}$ The upper limits were calculated using Gaussians of
zero width to model features narrower than the RGS resolution.  Limits
on the strength of O lines are not constraining due to poor statistics
at low energies.  These upper limits are similar to or below the
fluxes measured from the lines detected in 4U 1626--67, indicating a
constraining sensitivity was achieved in this {\it XMM-Newton} observation.}
\vspace{-1.0\baselineskip}
\end{footnotesize}
\end{table}


\begin{references}

\reference{} Arnaud, K. A., 1996, Astronomical Data Analysis Software
and Systems V, eds. G. Jacoby and J. Barnes, p17, ASP Conf. Series
vol. 101

\reference{} Asai, K., Dotani, T., Nagase, F., \& Mitusda, K., 2000,
ApJS, 131, 571

\reference{} Bhattacharya, D., \& van en Heuvel, E. P. J., 1991, Phys
Rep., 203, 1

\reference{} Bildsten, L., 2002, ApJ, in press, astro-ph/0208164

\reference{} Bildsten, L., \& Chakrabarty, D., 2002, ApJ, 557, 292

\reference{} Campana, S., et al., 2002, ApJ, in press,
astro-ph/0206376

\reference{} Chakrabarty, D., 1998, ApJ, 492, 342

\reference{} Chakrabarty, D., \& Morgan, E. H., 1998, Nature, 394, 346

\reference{} Ehle, M., et al., 2002, IAU Circ. 7872

\reference{} Galloway, D. K., Chakrabarty, D., Morgan, E., \&
Remillard, R., 2002, ApJ, in press, astro-ph/0206493

\reference{} Gierlinski, M., Done, C., \& Barret, D., 2002, MNRAS,
331, 141

\reference{} Heindl, W. A., \& Smith, D. A., 1998, ApJ, 560, 35L

\reference{} den Herder, J. W., 2002, RGS-SRON-RP-CAL-01/006,
available at http://xmm.vilspa.esa.es/ccf/documents/\# general

\reference{} Juett, A. M., Psaltis, D., \& Chakrabarty, D., 2001, ApJ,
560, L59

\reference{} Juett, A. M., Galloway, D. K., \& Chakrabarty, D., 2002,
ApJ, submitted, astro-ph/0208543

\reference{} Kallman, T. R., Angelini, L., Boroson, B., \& Cottam, J.,
2002, ApJ, subm., astro-ph/0209010

\reference{} Kirsch, M., 2002, XMM-SOC-CAL-TN-0018, available at
http://xmm.vilspa.esa.es/ccf/documents/\# general

\reference{} Lewin, W. H. G., van Paradijs, J., \& Taam R. E., 1993,
SSRv, 62, 223

\reference{} Markwardt, C. B., \& Swank, J. H., 2002a, IAU Circ. 7867

\reference{} Markwardt, C. B., \& Swank, J. H., 2002a, IAU Circ. 7870

\reference{} Markwardt, C. B., Swank, J. H., Strohmayer, T. E., in~'t
Zand, J. J. M, \& Marshall, F. E., 2002, ApJ, 575, L21

\reference{} Miller, J. M., et al., 2002a, ATEL 90

\reference{} Miller, J. M., et al., 2002b, ATEL 91

\reference{} Miller, J. M., et al., 2002c, ApJ, 570, L69

\reference{} Morrison, R., \& McCammon, D., 1983, ApJ, 270, 119

\reference{} Orlandini, M., et al., 1998, ApJ, 500, 163L

\reference{} Remillard, R. A., Swank, J., \& Strohmayer, T., 2002, IAU
Circ. 7893

\reference{} Schulz, N. S., Chakrabarty, D., Marshall, H. L.,
Canizares, C. R., Lee, J. C., \& Houck, J., 2001, ApJ, 563, 941
 
\reference{} Stella, L, White, N. E., \& Priedhorsky, W., 1987, ApJ, 312, L17

\reference{} Strohmayer, T. E., 2001, AdSpR, 28, 511

\reference{} Titarchuk, L., 1994, ApJ, 434, 313

\reference{} Verner, D. A., Yakovlev, D. G., Band, I. M., \&
Trzhaskovskaya, M. B., 1993, ADNDT, 55, 233

\reference{} White, N. E., Swank, J. H., \& Holt, S. S., 1983, ApJ,
270, 711

\reference{} Wilms, J., Nowak, M. A., Dove, J. B., Fender, R. P., \&
Di Matteo, T., 1999, ApJ, 522, 460

\reference{} Wijnands, R., \& van der Klis, M., 1998a, Nature, 394, 344

\reference{} Wijnands, R., \& van der Klis, M., 1998b, ApJ, 507, 63L

\reference{} Yungelson, L. R., Nelemans, G., \& van den Heuvel,
E. P. J., 2002, A \& A, 388, 546

\end{references}
\end{document}